\documentclass[11pt,twoside]{article}  
\usepackage{apn3conf}

\begin{document}   

%
%
%
%

\title{Mid-Infrared Imaging and Modelling of the Dust Shell around Post-AGB star HD 187885 (IRAS 19500-1709)}
\titlemark{Mid-Infrared Imaging and Modelling of IRAS 19500-1709}

%
%
%

\author{K.L. Clube, T.M. Gledhill}
\affil{Dept. of Physics, Astronomy \& Maths, University of Hertfordshire, College Lane, Hatfield, UK, AL10 9AB}

%
%

\contact{Kim Clube}
\email{kclube@star.herts.ac.uk}

%
%
%
%
%

\paindex{Clube, K. L.}
\aindex{Gledhill, T. M.}     

%
%

\authormark{Clube \& Gledhill}

%
%

\keywords{IRAS 19500-1709, HD 187885, radiative transfer, post-AGB, Mid-IR}


\begin{abstract}          
We present 10 and 20$\mu$m images of IRAS 19500-1709 taken with the mid-infrared camera, OSCIR, mounted on the Gemini North Telescope. We use a 2-D dust radiation transport code to fit the spectral energy distribution from UV to sub-mm wavelengths and to simulate the images.
\end{abstract}

%
%

\section{Introduction}
The circumstellar envelopes (CSEs) of post-AGB stars are cool and dust rich and hence radiate most of their energy at mid-infrared wavelengths. IRAS 19500-1709 is associated with the high galactic latitude F2-3I (Parthasarathy, Pottasch \& Wamsteker 1988) post-AGB star HD 187885 and has the double-peaked spectral energy distribution (SED) typical of post-AGB stars with detached CSEs (Hrivnak, Kwok \& Volk 1989). The expansion velocity of the envelope, based on CO-line emission, is 11 km s$^{-1}$ with wings up to 30 km s$^{-1}$ (Likkel et al. 1987). The envelope is carbon rich, showing no OH or H$_{2}$O maser emission (Likkel 1989). It has a broad emission feature from 10-13$\mu$m peaking at 12$\mu$m which could be attributed to a polycyclic aromatic hydrocarbon (PAH), such as chrysene (Justtanont et al. 1996). It has a weak 21$\mu$m feature (Justtanont et al. 1996) and a broad feature around 30$\mu$m which has been modelled with non-spherical magnesium sulphide (MgS) dust grains (Hony, Waters \& Tielens 2002). Imaging polarimetry at near-infrared wavelengths shows that IRAS 19500-1709 has a bipolar structure in scattered light (Gledhill et al. 2001).
 We present mid-infrared imaging of IRAS 19500-1709 using the OSCIR camera mounted on the 8.1-m Gemini North Telescope which provides information on the inner part of the CSE. We fit the spectral energy distribution (SED) from UV to sub-mm wavelengths and simulate the images using a 2-D dust radiation transport (RT) code in order to derive the physical and chemical properties of the dust in the CSE.

\section{Observations}
Using the OSCIR camera on Gemini North, we obtained 10 and 20$\mu$m (N and Q3 band) images of IRAS 19500-1709 with spatial resolution of 0.5 arcsecs (N band) to gain information on the inner part of the CSE (Fig.1). The object is extended relative to the point spread function (PSF).

\begin{figure}
\plottwo{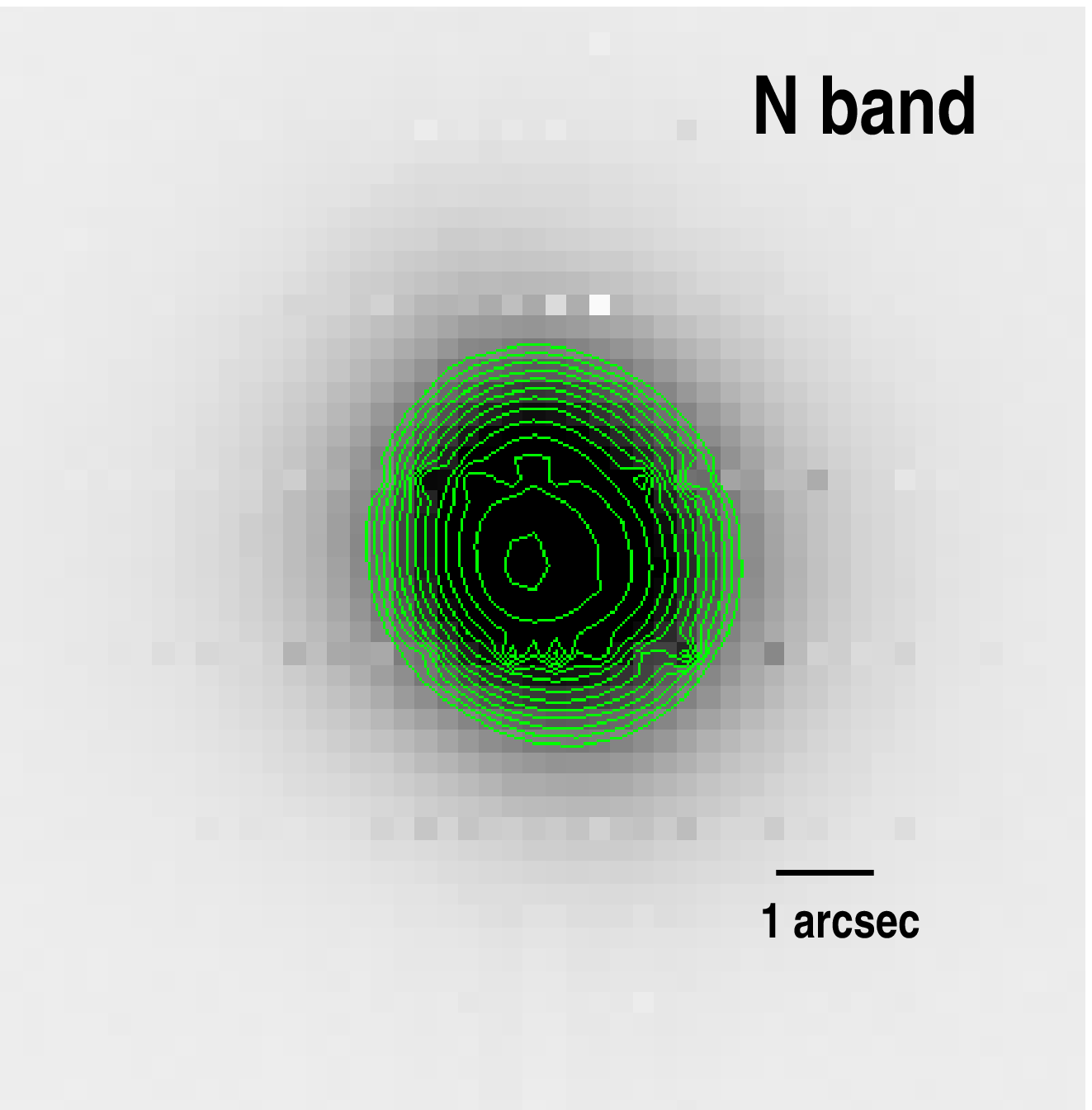}{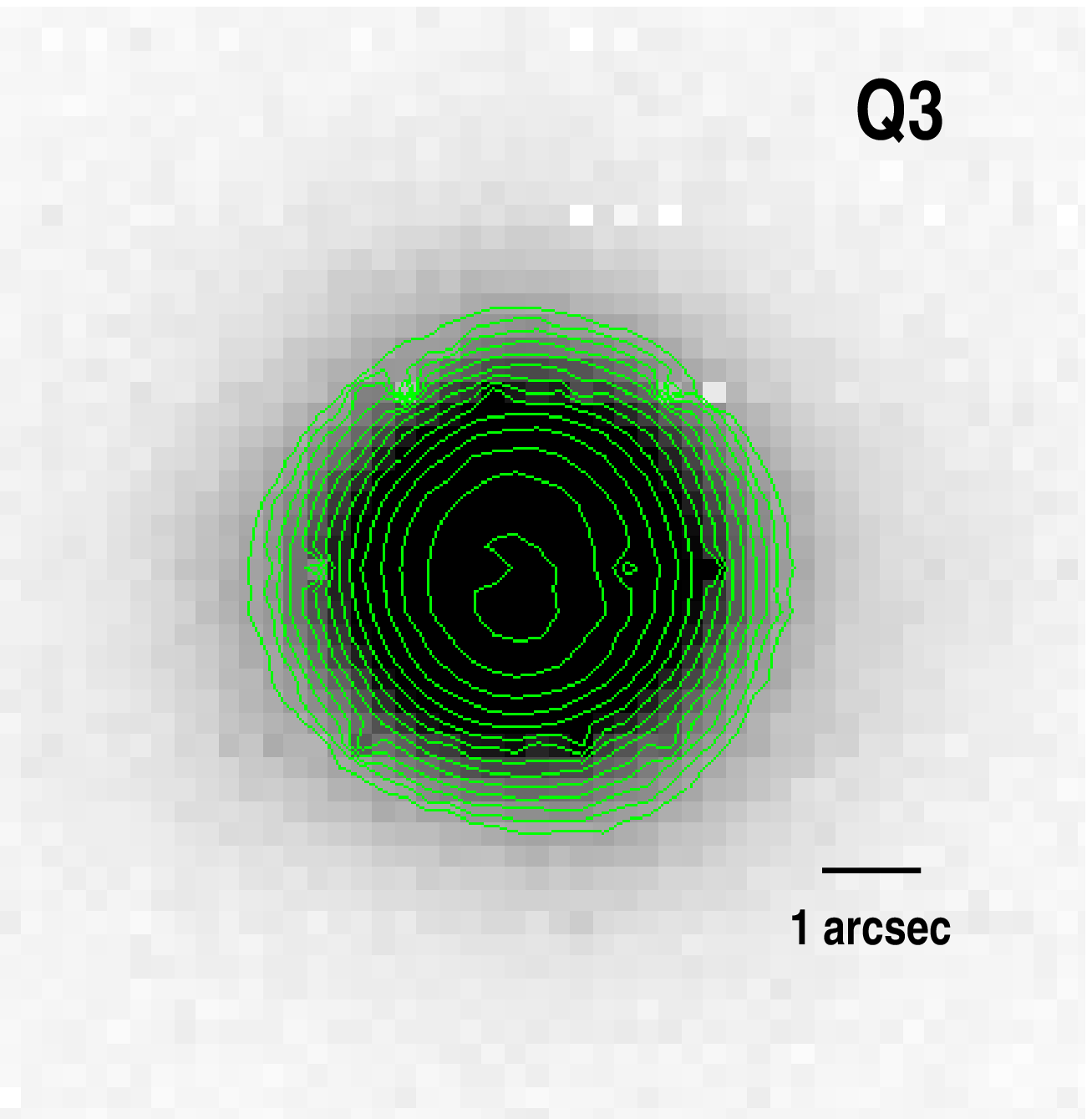}
\caption{Images of IRAS 19500-1709 from the OSCIR camera on Gemini North using N and Q3 filters.}
\end{figure}

 The N band suggests the object is elongated in a N-S direction. The outer contours appear rounder in the Q band image but there is evidence of elongation in the centre in a N-S direction.. Also the brightness peak in the N-band image appears approximately 0.3 arcsecs off centre to the east, relative to the outer contours.

\section{Modelling of the Dust Shell}
We modelled the images and SED of IRAS 19500-1709 using a 2-D axisymmetric RT dust code (Efstathiou \& Rowan-Robinson 1990) to derive the physical and chemical properties of dust in the envelope. The SED is constrained by the available observational data, for IRAS 19500-1709, taken from the literature which includes a 7.6-23.6$\mu$m UKIRT spectrum (CGS3) provided by K.Justannont, optical and near-IR photometry (Hrivnak, Kwok \& Volk 1989), 8.5, 10 and 12.2$\mu$m flux (Meixner et al. 1997), IRAS fluxes, IUE data\footnote{Based on INES data from the IUE satellite}, ISO SWS\footnote{ISO SWS06 observing program rszczerba-PPN30} and LWS\footnote{ISO LWS01 observing program mbarlow-dust 3} spectra and SCUBA flux (Van der Veen et al. 1994). Fig. 2 shows our best fit to date.

\begin{figure}
\epsscale{0.7}
\plotone{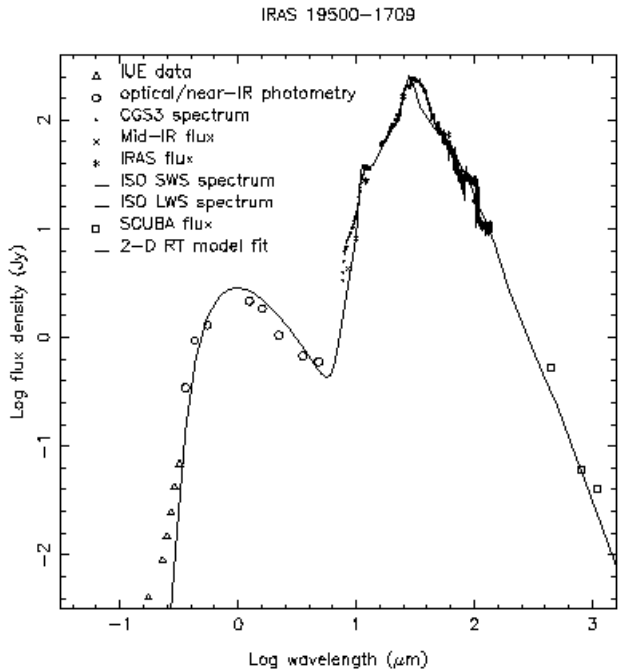}
\caption{SED and model fit (solid black line) for 19500-1709.}

\end{figure}

 In this model we use a spherically symmetric shell with dust composed of 0.01-2$\mu$m amorphous carbon (amC), silicon carbide (SiC) and MgS grains. The amC dust fits the continuum of the SED quite well. We include SiC to account for the presence of the infrared emission feature from 10-13$\mu$m. We attempted to fit this feature using PAHs but found that they produced emission features in the 16-25$\mu$m region of the spectrum which are not seen in the data. The MgS is included as being the possible carrier of the 30$\mu$m band, although it is possible that this feature is due to an unknown carbonaceous component. We used optical constants for Mg$_{x}$Fe$_{(1-x)}$S from Begemann et al ((1994) with x=0.9 as an approximation to MgS (also see Hony et al. 2003). These data do not extend below 10$\mu$m however, so it was necessary to approximate the optical and UV extinction cross-sections to ensure that the MgS grains absorb sufficient radiation at short wavelengths to radiate in the mid-IR. This was done by adopting the optical constants of graphite below 4$\mu$m and interpolating between 4 and 10$\mu$m to produce a smooth extinction curve.  We assume all dust grains are spherical with a size distribution of n(a)$\alpha$a$^{-q}$ with q=5. The dust is composed of 30\% amC (temp. 87-167K), 40\% SiC (temp. 79-150K) and 30\% MgS (temp. 59-121K). The central star is assumed to be a black body with T=7500K. The ratio of the stellar radius to the inner radius of the shell is determined to be 9x10$^{-5}$ in our model and A$_{v}$=1.5. The ratio of the inner shell radius to the outer shell radius is 0.2 which produces sufficient far-IR emission to fit the data longward of 100$\mu$m.  Extending the outer boundary any further than this produces too much far-IR emission. 

\begin{figure}
\epsscale{0.5}
\plottwo{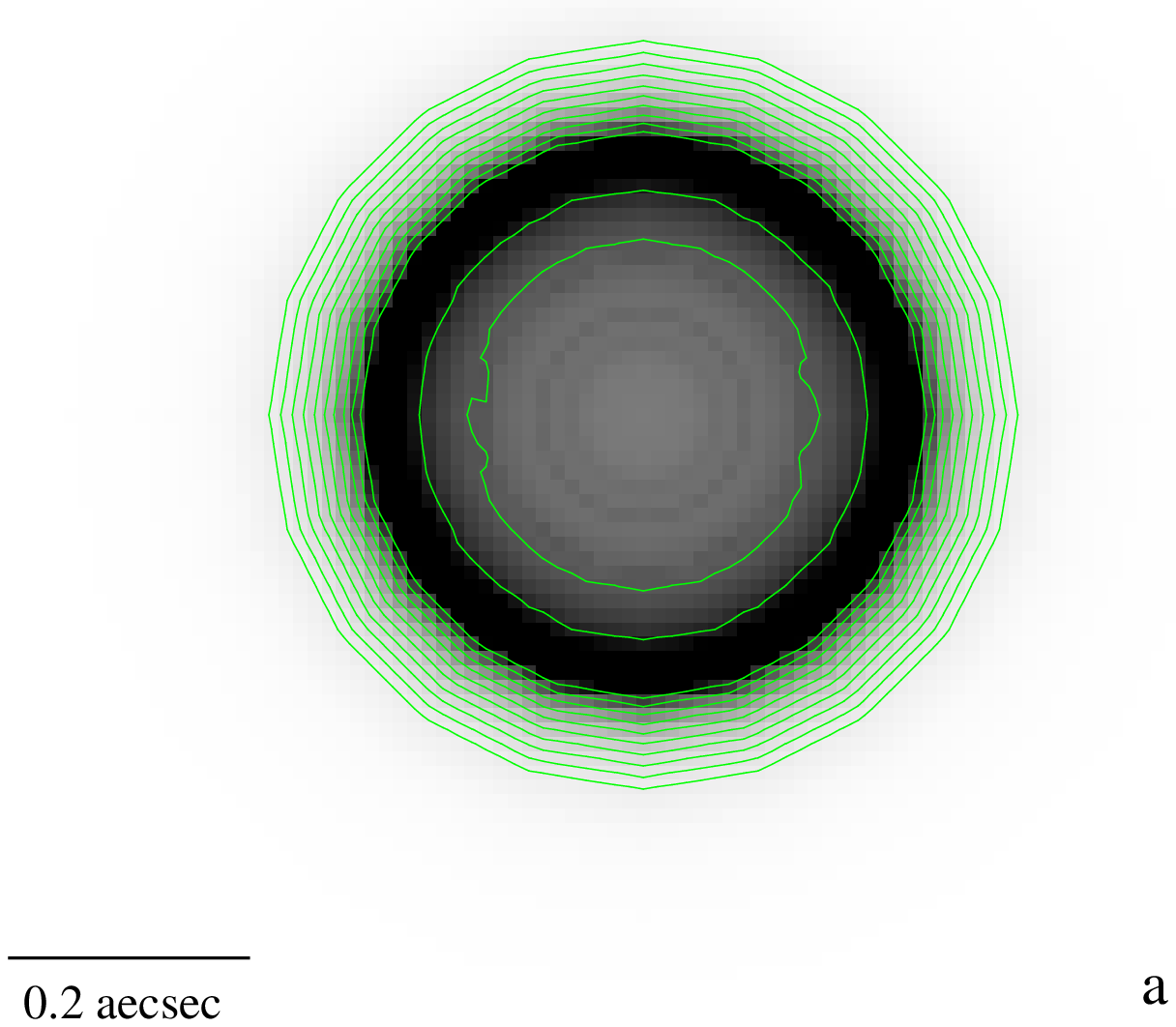}{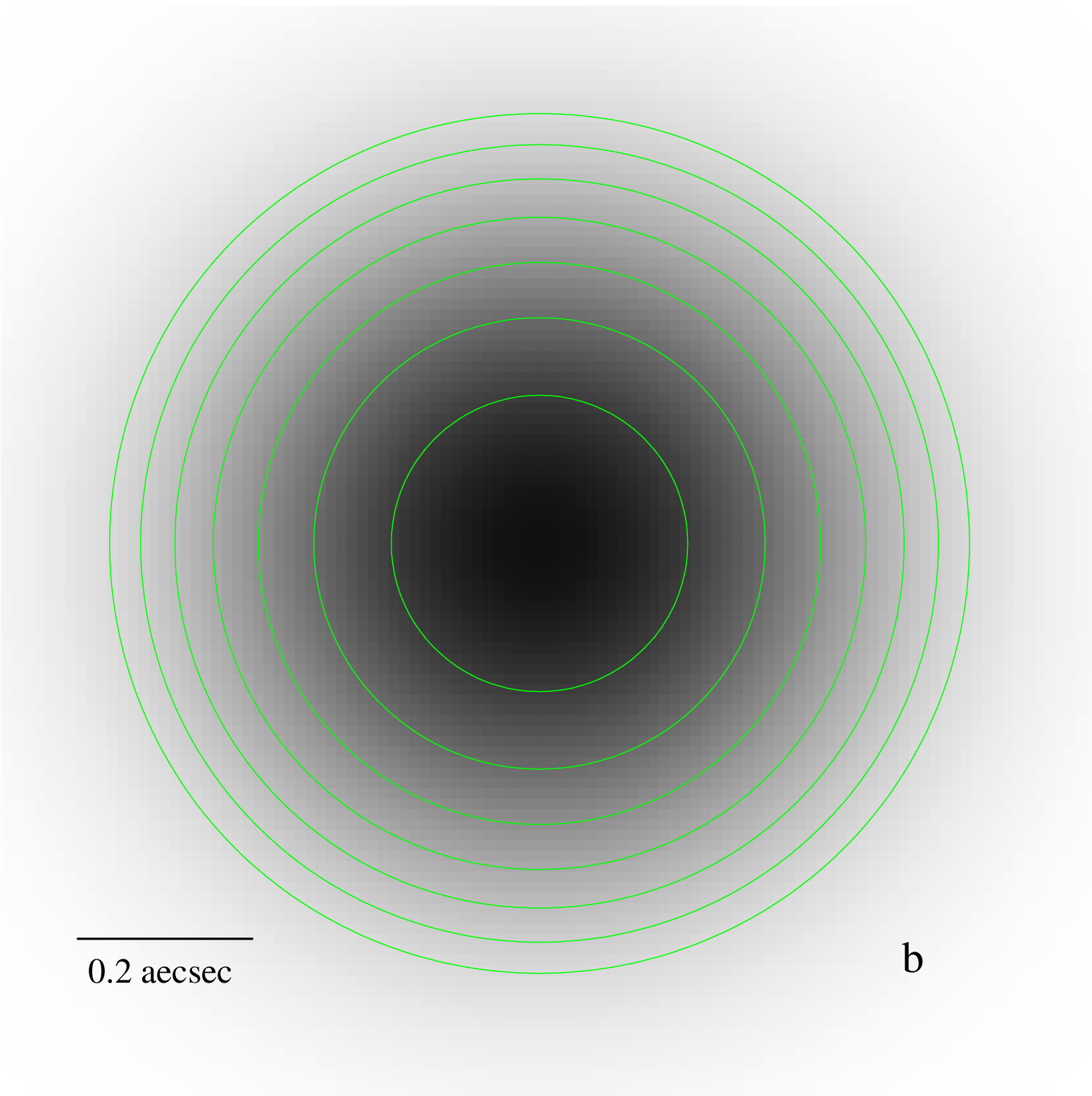}
\caption{a Raw N-band model image. b Smoothed N-band model image. The images are greyscaled such that the peak value of the smoothed image is 50\% that of the raw image. The contours of both images are logarithmically spaced in steps of 0.3 magnitudes. The lowest contour is the same in both cases.}
\end{figure} 

In Fig.3a we show a raw model image corresponding to the best fit envelope parameters. In Fig.3b the model image is smoothed using a gaussian filter to simulate the OSCIR N-band image. The degree of smoothing (FWHM of the Gaussian) is determined by matching profiles between the model and OSCIR images. A gaussian smoothing function with a FWHM of 40 pixels provides the best match. This corresponds to a model pixel size of 0.01 arcsec. Using this procedure we estimate that the inner radius of the detached shell is 0.2 arcsec.

\section{Future Work}
 There appears to be excess emission at 7-10$\mu$m as well as the broad feature around 30$\mu$m. We have attempted to fit this but the MgS peak is too narrow. This may be due to the MgS grains being too cold or because we have used spherical grains. Variations in grain shape can influence the emission profile with elliptical grains causing a wider peak (Hony, Waters \& Tielens 2002). We will be running axisymmetric models to determine whether we can reproduce the elongation seen in the OSCIR images.


\begin{references}
\reference Begemann, B.\ et al.\ 1994, \apj, 423, L71
\reference Efstathiou, A; \& Rowan-Robinson, M.\ 1990, \mnras, 245, 275
\reference Gledhill, T.\ M.\ et al.\ 2001, \mnras, 322, 321
\reference Hony, S; Waters, L.\ B.\ F.\ M; \& Tielens, A.\ G.\ G.\ M.\ 2002, \aap, 390, 533
\reference Hony, S.\ et al.\ 2003, \aap, 402, 211
\reference Hrivnak, B.\ J; Kwok, S; \& Volk, K.\ M.\ 1989, \apj, 346, 265
\reference Justtanont, K.\ et al.\ 1996, \aap, 309, 612
\reference Likkel, L.\ et al.\ 1987, \aap, 173, L11
\reference Likkel, L.\ 1989, \apj, 344, 350
\reference Meixner, M.\ et al.\ 1997, \apj, 482, 897
\reference Parthasarathy, M; Pottasch, S.\ R; \& Wamsteker, W.\ 1988, \aap, 203, 117
\reference Van der Veen, W.\ E.\ C.\ J.\ et al.\ 1994, \aap, 285, 551
\end{references}
\end{document}